\documentclass[aps,prb,twocolumn,groupedaddress,showpacs,amsmath,amssymb]{revtex4}

\usepackage{graphicx}
\usepackage{dcolumn}
\usepackage{bm}

\begin{document}

\title{Observations of two-dimensional quantum oscillations and ambipolar transport
in the topological insulator Bi$_{2}$Se$_{3}$ achieved by Cd doping}

\author{Zhi Ren}
\author{A. A. Taskin}
\author{Satoshi Sasaki}
\author{Kouji Segawa}
\author{Yoichi Ando}

\affiliation{Institute of Scientific and Industrial Research,
Osaka University, Ibaraki, Osaka 567-0047, Japan}

\date{\today}

\begin{abstract}
We present a defect-engineering strategy to optimize the transport
properties of the topological insulator Bi$_{2}$Se$_{3}$ to show a high
bulk resistivity and clear quantum oscillations. Starting with a
$p$-type Bi$_{2}$Se$_{3}$ obtained by combining Cd doping and a Se-rich
crystal-growth condition, we were able to observe a $p$-to-$n$-type
conversion upon gradually increasing the Se vacancies by post annealing.
With the optimal annealing condition where a high level of compensation
is achieved, the resistivity exceeds 0.5 $\Omega$cm at 1.8 K
and we observed two-dimensional Shubnikov-de Haas oscillations composed
of multiple frequencies in magnetic fields below 14 T.
\end{abstract}


%

\maketitle
\section{Introduction}
The three-dimensional (3D) topological insulator (TI) realizes a novel
quantum state of matter where a nontrivial $Z_2$ topology of the wavefunction
of the bulk valence band leads to the emergence of a ``topological"
surface state consisting of helically spin-polarized Dirac
fermions.\cite{K2,MB,Roy} The peculiar spin texture of the surface state
holds promise for novel spintronics and fault-tolerant topological
quantum computing, so there is a rush of research to address this
surface state.\cite{Kane,ZhangSC} However, most of the known TI
materials are poorly insulating in the bulk, making it difficult to
probe the surface state by transport experiments. For example,
Bi$_{2}$Se$_{3}$
is considered to be a promising TI material because it
has a relatively large ($\sim$0.3 eV) bulk band
gap and a nearly perfect Dirac cone as its topological
surface state;\cite{ZhangH,Xia}
however, no matter whether it is in the form of bulk
crystal,\cite{Xia,Eto} nanoribbon,\cite{Peng} or epitaxial thin
film,\cite{Xue} Bi$_{2}$Se$_{3}$ always accompanies a lot of Se
vacancies (usually $\sim$10$^{19}$ cm$^{-3}$) that act as electron
donors, and as a result, the residual bulk carriers hinder the transport
studies of the surface state of this material.

To achieve a bulk-insulating state in Bi$_{2}$Se$_{3}$, doping holes to
compensate for the residual electrons is a viable strategy. While this
was done through low-level substitution of Ca$^{2+}$ for
Bi$^{3+}$,\cite{Hor} the resulting disorder was so strong that no
Shubnikov-de Haas (SdH) oscillation from the surface state was
observed\cite{Checkelsky} in Bi$_{2-x}$Ca$_{x}$Se$_{3}$. A different
strategy was to partially substitute Sb for Bi, which apparently reduces
the Se vacancies; indeed, with a relatively large ($\sim$12\%) Sb
substitution, surface SdH oscillations were successfully observed in
$n$-type Bi$_{2-x}$Sb$_{x}$Se$_{3}$, but a very high magnetic field
($\sim$60 T) was required for the observation.\cite{Analytis1} It is to
be noted that with the Sb doping one can never cross the band gap to
reach the $p$-type regime, and hence the tuning of the chemical
potential to the Dirac point is impossible. This is a pity, because
Bi$_{2}$Se$_{3}$ is attractive for its isolation of the Dirac point from
the bulk bands. Therefore, it is desirable to find a suitable $p$-type
dopant to access the Dirac point while keeping the mobility to be
sufficiently high for the surface state to be studied by the SdH
oscillations.

In this paper, we show that tactful defect engineering in
Bi$_{2}$Se$_{3}$ employing Cd doping in combination with Se-vacancy
tuning provides a useful means to control the chemical potential across
the band gap. In the literature,\cite{Horak} whereas Cd in
Bi$_{2}$Se$_{3}$ was shown to behave as an acceptor, Cd-doped
Bi$_{2}$Se$_{3}$ crystals always remained $n$-type due to the low
solubility of Cd atoms in Bi$_{2}$Se$_{3}$; however, it has been
elucidated\cite{kohler,Analytis2,Butch} that increasing the Se content
in the Bi-Se melt for the crystal growth can suppress the formation of
Se vacancies and greatly reduce the residual bulk carrier density to the
level of $\sim$10$^{17}$ cm$^{-3}$. Therefore, even though the
solubility of Cd is low, one could achieve a $p$-type behavior by
combining Cd doping and a Se-rich growth condition. Actually, we
obtained a $p$-type sample with this strategy and, furthermore, starting from
the $p$-type sample, we could gradually increase the Se vacancies by
careful post-annealing and achieve a high level of compensation, at
which the sample becomes optimally bulk-insulating and presents
two-dimensional (2D) SdH oscillations below 14 T.

\section{Experimental Details}

The single crystals of Cd-doped Bi$_{2}$Se$_{3}$ were grown by using
elemental shots of Bi (99.9999\%), Cd (99.99\%) and Se (99.999\%) as
starting materials. To maximize the Cd content in the crystal, excess Cd
and a mixture of Bi and Se with a ratio of Bi:Se = 32:68 were melted in
a sealed evacuated quartz tube at 750 $^{\circ}$C for 48 h with
intermittent shaking to ensure a homogeneity, followed by cooling slowly
to 550 $^{\circ}$C and then annealing at the same temperature for one
week. The resulting crystals are easily cleaved along the basal plane,
revealing a silvery mirror-like surface. The X-ray diffraction
measurements confirmed the crystal to be single phase with the proper
Bi$_{2}$Se$_{3}$ structure. The actual Cd content was determined by the
inductively-coupled plasma atomic-emission spectroscopy (ICP-AES) to be
0.0020(2). The as-grown crystals were examined by X-ray Laue analysis
and cut into single-domain, thin bar-shaped samples with typical
dimensions of 3$\times$1$\times$0.2 mm$^{3}$.

For each annealing
experiment, samples weighing about 4.2 mg were sealed in evacuated
quartz tubes and annealed at a given temperature for one week, followed
by quenching into cold water. All the samples used in this work were
taken from the same part of the same batch, and the variation of the Cd
content was confirmed to be negligible by the ICP-AES analysis. To avoid
possible surface contamination, the surface layer of the annealed
crystals were removed using adhesive tapes before transport
measurements.

It worth mentioning that in our annealing experiments, we took
precautions to minimize the uncertainty in the annealing temperature.
For each annealing run, we placed the quartz tube at the same position
of the same furnace so that the temperature gradients in
the furnace as well as the thermocouple calibration errors do not affect the
annealing result. Also, the environment temperature was kept constant
during this experiment to minimize the temperature fluctuations between
different annealing runs. As a result, the annealing temperature $T_{\rm
anneal}$ was very reproducible and its variation between different runs
with nominally the same $T_{\rm anneal}$ was within $\pm$1 $^{\circ}$C.

The in-plane resistivity $\rho_{xx}$ and the Hall coefficient $R_{\rm
H}$ were measured in a Quantum Design Physical Properties Measurement
System (PPMS-9) down to 1.8 K, for which the electrical contacts were
prepared by using room-temperature-cured silver paste. In addition, one
of the high-resistivity samples was brought to a 14-T magnet for
detailed SdH-oscillation measurements using an ac six-probe method, in
which four lock-in amplifiers were employed to measure both the primary
and the second-harmonic signals in the longitudinal and transverse
channels at a frequency of 19 Hz. The SdH-oscillation data were taken by
sweeping the magnetic fields between $\pm$14 T with the rate of 0.3
T/min, during which the temperature was stabilized to within $\pm$5 mK.

\section{Results and Discussions}

\subsection{$p$-type Bi$_{2-x}$Cd$_{x}$Se$_{3}$}

\begin{figure}
\includegraphics*[width=7.5cm]{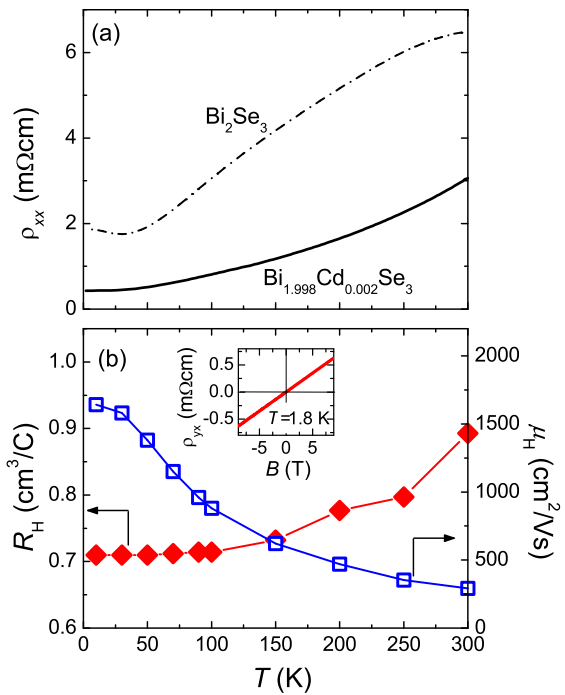}
\caption{(Color online)
(a) Temperature dependences of $\rho_{xx}$ of an as-grown
Bi$_{1.998}$Cd$_{0.002}$Se$_{3}$ crystal (solid line) and a pristine
Bi$_{2}$Se$_{3}$ crystal (dash-dotted line) grown in the same Se-rich
condition. The pristine sample is $n$-type with $n_{\rm e} \sim$
7$\times$10$^{17}$ cm$^{-3}$, while the Cd-doped sample is $p$-type with
$n_{\rm h} \sim$ 9$\times$10$^{18}$ cm$^{-3}$. Note that the
low-temperature resistivity upturn in the pristine sample is absent in
the Cd-doped sample. (b) Temperature dependences of $R_{\rm H}$
(left axis) and $\mu_{\rm H}$ for the Cd-doped sample; inset shows
the magnetic field dependence of $\rho_{yx}$ in this sample at 1.8 K.
}
\label{fig1}
\end{figure}

The temperature dependence of the in-plane resistivity $\rho_{xx}$ of a
Bi$_{1.998}$Cd$_{0.002}$Se$_{3}$ crystal grown in the Se-rich condition
is shown in Fig. 1(a), together with the data for a pristine
Bi$_{2}$Se$_{3}$ sample grown with the same Bi/Se ratio. The pristine
Bi$_{2}$Se$_{3}$ crystal shows an essentially metallic behavior with a
weak resistivity upturn below $\sim$30 K, which is typical for
low-carrier-density Bi$_{2}$Se$_{3}$; \cite{kohler,Analytis2,Butch}
indeed, the Hall coefficient $R_{\rm H}$ in this sample at 1.8 K
corresponds to the bulk electron density $n_{\rm
e}$$\sim$7$\times$10$^{17}$ cm$^{-3}$, which is very small for
Bi$_{2}$Se$_{3}$. On the other hand, in Bi$_{1.998}$Cd$_{0.002}$Se$_{3}$
the resistivity upturn is absent and the $\rho_{xx}$ value is lower,
suggesting a higher carrier density. In fact, as shown in Fig. 1(b),
$R_{\rm H}$ in the Bi$_{1.998}$Cd$_{0.002}$Se$_{3}$ crystal is positive
and its value at 1.8 K corresponds to the hole density $n_{\rm
h}$$\sim$9$\times$10$^{18}$ cm$^{-3}$, implying that the Cd doping has
created $\sim$9.7$\times$10$^{18}$ cm$^{-3}$ of holes ($\sim$0.8 hole
per Cd atom) that outnumbers the electrons coming from Se
vacancies.\cite{Horak} The reason for the smaller number of doped holes
compared to the Cd concentration is most likely that a small portion of
Cd atoms occupy the interstitial site and act as donors. In Fig. 1(b),
$R_{\rm H}$ is only weakly dependent on temperature and the Hall
resistivity $\rho_{yx}$ is perfectly linear in $B$ (as shown in the
inset), reflecting the metallic nature of the as-grown
Bi$_{1.998}$Cd$_{0.002}$Se$_{3}$ sample which is due to a single type of
carriers ({\it i.e.}, the bulk holes). The Hall mobility $\mu_{\rm H}$
(= $R_{\rm H}$/$\rho_{xx}$), also shown in Fig. 1(b), increases with
decreasing temperature, reaching $\sim$1600 cm$^{2}$/Vs at 1.8 K.

\subsection{$p$-to-$n$-type conversion by post annealing}

\begin{figure}
\includegraphics*[width=8.5cm]{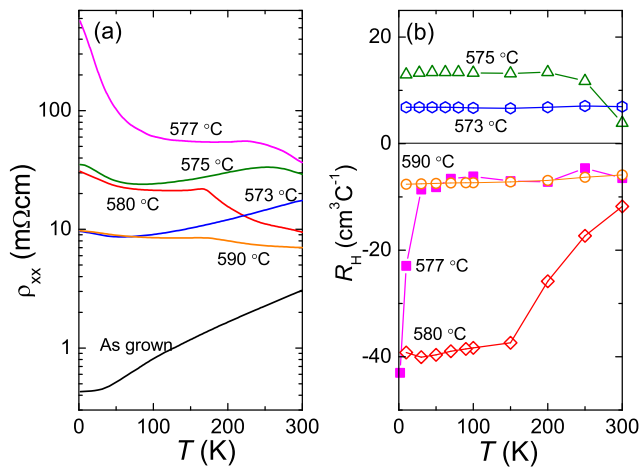}
\caption{(Color online)
(a) Temperature dependences of $\rho_{xx}$ for
Bi$_{1.998}$Cd$_{0.002}$Se$_{3}$ crystals annealed at different
temperatures in evacuated quartz tubes. The data for the as-grown
crystal is also shown for comparison. The low-temperature resistivity
values span three orders of magnitude. (b) Temperature dependences of
$R_H$ for the same series of samples. As the annealing temperature is
increased, the crystal becomes less $p$-type and is eventually converted
to $n$-type. This conversion occurs with the annealing temperature of
$\sim$577 $^{\circ}$C.
}
\label{fig2}
\end{figure}

Annealing the as-grown Bi$_{1.998}$Cd$_{0.002}$Se$_{3}$ crystals in
evacuated quartz tubes has a drastic effect on its transport properties.
Figure 2(a) shows how the temperature dependence of $\rho_{xx}$ changes
upon annealing in a narrow temperature window between 573 and 590
$^{\circ}$C. One can see that $\rho_{xx}$($T$) evolves {\it
nonmonotonically} with the annealing temperature, $T_{\rm anneal}$;
namely, $\rho_{xx}$ initially increases with $T_{\rm anneal}$ until
$T_{\rm anneal}$ exceeds 577 $^{\circ}$C, after which $\rho_{xx}$
decreases as $T_{\rm anneal}$ is further increased. Notably, $\rho_{xx}$
of the sample annealed at 577 $^{\circ}$C show a high value of 0.5
$\Omega$cm at 1.8 K, which is three orders of magnitude larger than that
of the as-grown sample. Figure 2(b) shows the typical temperature
dependences of low-field $R_{\rm H}$ data (defined as $R_{\rm H}$ =
$\rho_{yx}$/$B$ for $B \approx$ 0) for different $T_{\rm anneal}$, which
indicates that more and more electron carriers are introduced as $T_{\rm
anneal}$ is increased, and the sign change from $p$-type to $n$-type
occurs around $T_{\rm anneal}$ = 577 $^{\circ}$C.

\begin{figure}
\includegraphics*[width=8.5cm]{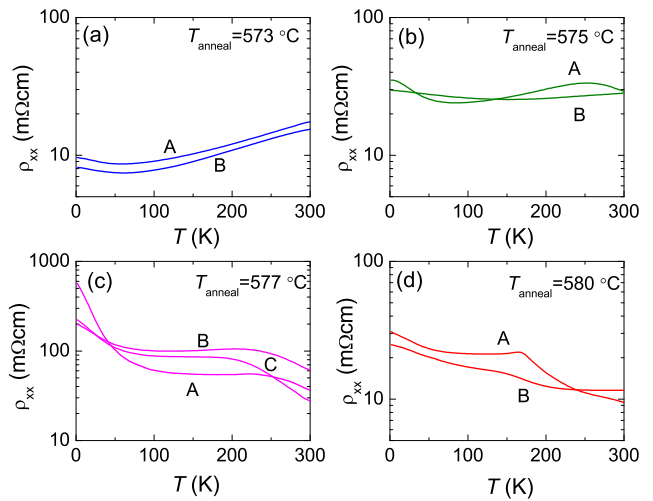}
\caption{(Color online)
(a-d) Reproducibility of the $\rho_{xx}(T)$ data in samples annealed at the
same temperature, demonstrated for four different values of
$T_{\rm anneal}$ indicated in each panel.
}
\label{fig3}
\end{figure}

\begin{figure*}[btp]
\includegraphics*[width=16cm]{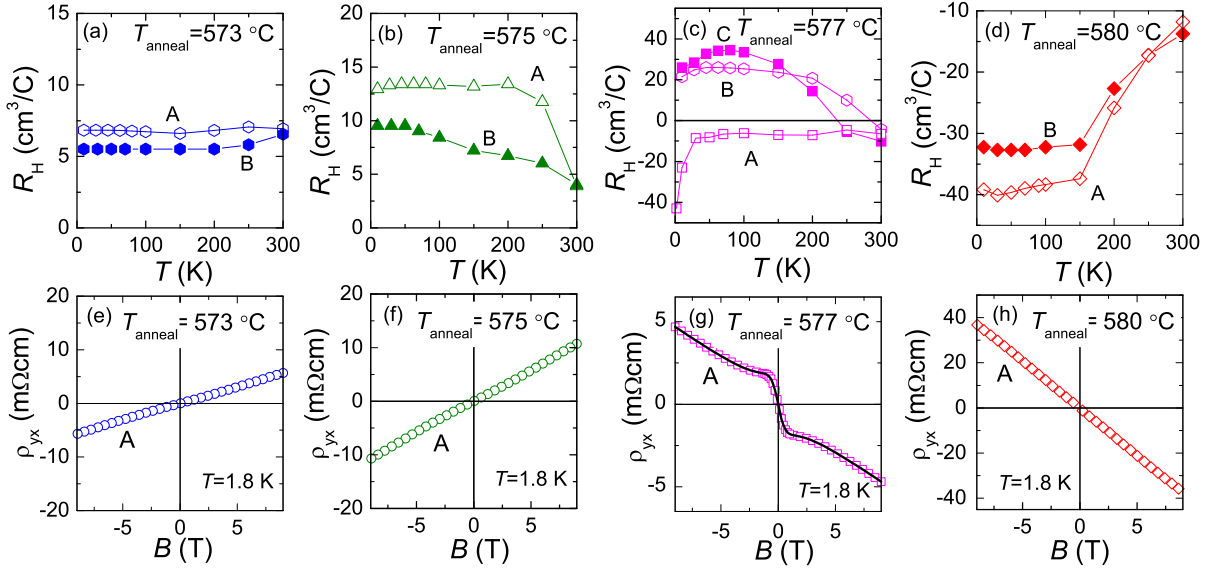}
\caption{(Color online)
(a-d) Reproducibility of the $R_{\rm H}(T)$ data in samples annealed at the
same temperature, demonstrated for four different values of
$T_{\rm anneal}$ indicated in each panel.
(e-f) Magnetic-field dependences of
$\rho_{yx}$ measured in sample ``A" of each $T_{\rm anneal}$.
The solid line in (g) is the two-band-model fitting to the data.
}
\label{fig4}
\end{figure*}

Since the drastic change in the transport properties occurs in a very
narrow temperature window (573 -- 580 $^{\circ}$C), one may wonder about the
reproducibility of the result. As a matter of fact, the observed change
was quite reproducible, as demonstrated in Figs. 3 and 4. Figures
3(a)-3(d) show the $\rho_{xx}(T)$ data for at least two samples annealed
at the same temperature, where one can see that the behavior for each
$T_{\rm anneal}$ is essentially reproducible. Figures 4(a)-4(d) show the
corresponding $R_{\rm H}(T)$ data for the same sets of samples; here,
except for the case of $T_{\rm anneal}$ = 577 $^{\circ}$C [Fig. 4(c)],
we observed reasonable reproducibility [Figs. 4(a), 4(b), and 4(d)]. For
$T_{\rm anneal}$ = 577 $^{\circ}$C, two of the three samples (A and B)
showed a sign change in $R_{\rm H}$ from negative to positive upon
lowering temperature from 300 K, whereas the $R_{\rm H}$ of sample C
remained negative in the whole temperature range. Actually, this
variation in the behavior of $R_{\rm H}$ indicates that the $T_{\rm
anneal}$ = 577 $^{\circ}$C samples are at the verge of the
$p$-to-$n$-type conversion.

Figures 4(e)-4(h) show the $\rho_{yx}(B)$ curves measured in sample ``A"  of
each $T_{\rm anneal}$. One can
clearly see that the curve in Fig. 4(g) for $T_{\rm anneal}$ = 577 $^{\circ}$C
is nonlinear, indicating that
there are at least two bands contributing to the transport. In
topological-insulator samples with a large bulk resistivity, this kind
of nonlinear $\rho_{yx}(B)$ curves are indications of the surface
channels making noticeable contributions.\cite{BTS_Rapid,Steinberg,Qu,Oh,BSTS_PRL}
Therefore, the consistently
high resistivity [Fig. 3(c)] together with the complex
behaviors of the Hall signal are likely to be a signature of a high level
of compensation achieved in the samples annealed at 577 $^{\circ}$C; in
other words, in those samples the acceptors and donors are nearly equal
in number and their delicate balance can easily change the sign of $R_{\rm
H}$. It is to be emphasized that our data demonstrate that this high
level of compensation is reproducibly achieved with $T_{\rm anneal}$ =
577 $^{\circ}$C. In samples annealed at other temperatures, the
$\rho_{yx}(B)$ behavior is almost linear [Figs. 4(e), 4(f), and 4(h)],
suggesting that the
contribution of the surface to the transport properties is minor.
In passing, the collection of $\rho_{xx}(T)$ and $R_{\rm H}(T)$ data
shown in Fig. 2 for varying $T_{\rm anneal}$ are for sample ``A" of
each $T_{\rm anneal}$ shown in Figs. 3 and 4.

\subsection{Defect chemistry}

The above observation that a drastic change in the transport properties
of Bi$_{1.998}$Cd$_{0.002}$Se$_{3}$ occurs in a very narrow temperature
window might seem surprising. However, this behavior can be readily
understood by examining the defect chemistry associated with the
annealing. In the present system, there are mainly two different types
of charged defects, the aliovalent substitutional defect
Cd$_{\texttt{Bi}}^{\prime}$ and the vacancy defect
$V_{\texttt{\texttt{Se}}}^{\bullet\bullet}$,\cite{Hor,Horak} the former
acts as an acceptor and the latter as a donor. Therefore, the effective
charge-carrier density is determined by their competition and can be
expressed as $n_{\rm eff}$ = $[{\rm Cd}_{\texttt{Bi}}^{\prime}] -
2[V_{\texttt{Se}}^{\bullet\bullet}]$, where positive (negative) $n_{\rm
eff}$ denotes the hole (electron) density. In an as-grown sample, the
Cd$_{\texttt{Bi}}^{\prime}$ defects are dominant and $n_{\rm eff}$ is
positive; accordingly, the chemical potential lies in the valence band.
When annealed in evacuated quartz tubes, a portion of selenium goes into
the gas phase Se$_{2}$ in equilibrium with the solid phase,\cite{cafaro}
resulting in the formation of more $V_{\texttt{Se}}^{\bullet\bullet}$
defects while leaving Cd$_{\texttt{Bi}}^{\prime}$ unaffected (because
the $T_{\rm anneal}$ employed in the present study is much lower than
the melting temperature of 710 $^{\circ}$C). The equilibrated vapor
pressure of Se$_{2}$ increases with increasing $T_{\rm anneal}$,
\cite{cafaro} creating more $V_{\texttt{Se}}^{\bullet\bullet}$ and
eventually changing $n_{\rm eff}$ from positive to negative.

To be more quantitative, one may assume that the increase in the
Se-vacancy concentration upon annealing,
$\Delta[V_{\texttt{Se}}^{\bullet\bullet}]$, is directly reflected in the
increase in the number of Se$_{2}$ molecules in the quartz tube, which
determines the Se$_{2}$ vapor pressure $P_{\rm Se_{\rm 2}}$; in constant
volume, one expects a linear relation between
$\Delta[V_{\texttt{Se}}^{\bullet\bullet}]$ and $P_{\rm Se_{\rm 2}}$
if the Se$_{2}$ vapor behaves as an ideal gas. According
to Ref. \onlinecite{cafaro}, the equilibrated Se$_{2}$ vapor pressure
$P_{\rm Se_{\rm 2}}$ of Bi$_2$Se$_3$ is related to the absolute
temperature $T$ via
\begin{equation}
\log P_{{\rm Se_{\rm 2}}}{\rm [atm]} = A - B / T {\rm [K]},
\end{equation}
where $A$ = 7.81$\pm$0.50 and $B$ = 10870$\pm$640 for the temperature
range of 527 to 627 $^{\circ}$C. From this $T$ dependence of $P_{\rm
Se_{\rm 2}}$, one can infer that
$\Delta[V_{\texttt{Se}}^{\bullet\bullet}]$ is very sensitive to the
change in $T_{\rm anneal}$: For example, changing $T_{\rm anneal}$ by
just 1 $^{\circ}$C near 577 $^{\circ}$C results in a variation of
$\sim$3.5\% in $\Delta$$[V_{\texttt{Se}}^{\bullet\bullet}]$; therefore,
the expected change in $\Delta$$[V_{\texttt{Se}}^{\bullet\bullet}]$ upon
changing $T_{\rm anneal}$ from 575 to 580 $^{\circ}$C is as much as
$\sim$18\%. It is thus expected that a sign change in $n_{\rm eff}$
occurs abruptly in
the vicinity of $T_{\rm anneal}$ = 577 $^{\circ}$C where $n_{\rm eff}
\approx$ 0 (namely, [${\rm Cd}_{\texttt{Bi}}^{\prime}$] $\approx$
[$V_{\texttt{Se}}^{\bullet\bullet}$]), and results in a drastic change
in the transport properties as we observed.

\subsection{Surface quantum oscillations}

\begin{figure}\includegraphics*[width=8.5cm]{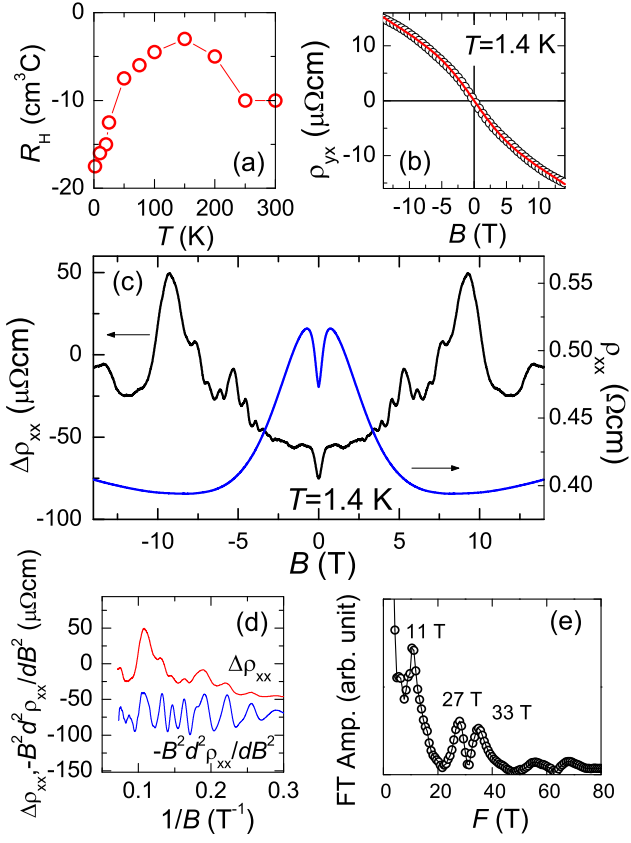}
\caption{(Color online)
(a) $R_{\rm H}(T)$ data of the sample C of $T_{\rm anneal}$ = 577
$^{\circ}$C after the Al$_2$O$_3$-coverage process. (b) $\rho_{yx}(B)$
data of the same sample measured at 1.4 K; the solid line is the result
of the two-band-model fitting. (c) Magnetic-field dependences of the
primary signal ($\rho_{xx}$) and the second-harmonic signal ($\Delta
\rho_{xx}$) measured at 1.4 K in magnetic fields along the $C_{3}$ axis.
(d) Plots of $\Delta\rho_{xx}(B)$ and
$d^2\rho_{xx}/dB^2$ vs the inverse magnetic field $1/B$. Clear SdH
oscillations can be seen in both data, and good agreements in the
positions of peaks and dips between the two curves are evident. (e) The
FT spectrum of $\Delta\rho_{xx}$ showing three prominent frequencies.
}
\label{fig5}
\end{figure}

From the above results, it is clear that the highest level of
compensation is achieved in samples annealed at 577 $^{\circ}$C. We
therefore measured the sample C of $T_{\rm anneal}$ = 577 $^{\circ}$C in
a 14-T magnet using a rotation sample holder to investigate its SdH
oscillations in detail. Before the high-field measurements, to protect
the surface state from aging, the top surface of the sample was covered
with Al$_{2}$O$_{3}$ in the following way: first, the crystal was
cleaved on both surfaces with adhesive tapes to reveal fresh surfaces,
mounted on a sample holder with GE varnish, and transferred into the
sputtering chamber; second, the top surface was cleaned by
bias-sputtering with Ar ions for 13 minutes and then, without breaking
the vacuum, a 540-nm-thick Al$_2$O$_3$ film was deposited by the rf
magnetron sputtering. After this process, gold wires were bounded to the
side faces by spot welding. Probably because of the sample heating
during the spot welding, the $\rho_{xx}$ value of this sample became
even larger than that shown in Fig. 3(c). Also, the sign of $R_{\rm H}$
at low temperature changed to negative after the process, showing the
Hall response similar to that of the sample A. The $R_{\rm H}(T)$
behavior of this sample C after the Al$_2$O$_3$-coverage process is
shown in Fig. 5(a), and its $\rho_{yx}(B)$ curve at 1.4 K is shown in
Fig. 5(b).

In our measurements using an ac lock-in technique, we simultaneously
recorded the primary and the second-harmonic signals during the
magnetic-field sweeps. Figure 5(c) shows the magnetic-field ($B$)
dependences of the primary signal ($\rho_{xx}$) and the second-harmonic
signal (denoted $\Delta \rho_{xx}$) measured at 1.4 K in magnetic fields along
the $C_{3}$ axis. One can see that the second-harmonic signal ($\Delta
\rho_{xx}$) shows pronounced oscillations, while in the primary signal
($\rho_{xx}$) the oscillations are hardly visible. To understand the
nature of the oscillations, we show in Fig. 5(d) the plots of $\Delta
\rho_{xx}$ and $d^2\rho_{xx}/dB^2$ (second derivative of the primary
signal) vs the inverse magnetic field $1/B$; a comparison between the
two curves indicates that they present essentially the same peak/dip
positions. While the waveforms are quite complicated, the Fourier
transform (FT) spectrum of $\Delta \rho_{xx}(B^{-1})$ shown in Fig. 5(e)
presents three well-defined peaks at $F_{1}$ = 11 T, $F_{2}$ = 27 T, and
$F_{3}$ = 33 T, indicating that the observed oscillations are SdH
oscillations with multiple frequencies.

Note that the second-harmonic in ac measurements is a distortion of the
input sine wave, and its occurrence is an indication of a nonlinear
response. In the present case, the SdH oscillations are apparently
giving rise to a peculiar non-ohmicity. This makes the second-harmonic
signal to be useful for observing the SdH oscillations with a high
sensitivity, although the detailed mechanism is not clear at the moment.

\begin{figure}\includegraphics*[width=8.5cm]{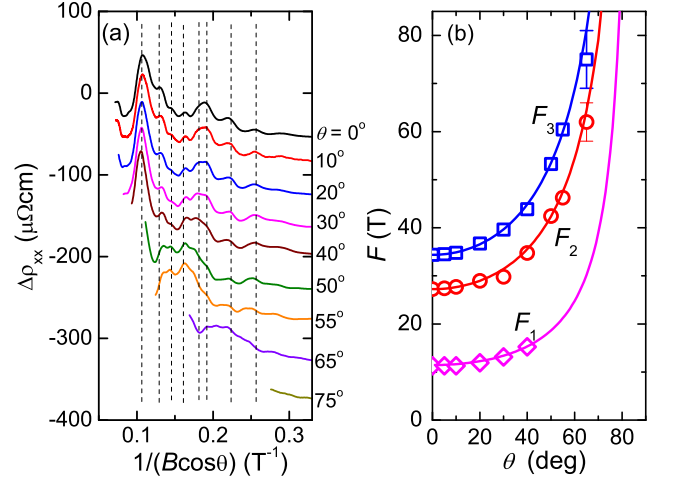}
\caption{(Color online)
(a) The $\Delta\rho_{xx}$ data for varying magnetic-field directions,
plotted as a function of $1/(B\cos\theta)$; dashed lines mark the
positions of the peaks.
(b) The three prominent frequencies in the FT spectra of the SdH oscillations
plotted as a function
of $\theta$. All the frequencies vary as $1/\cos\theta$, as indicated by
the solid lines.
}
\label{fig6}
\end{figure}

Figure 6(a) shows how the SdH oscillations observed in $\Delta\rho_{xx}$
change when the magnetic field is rotated, by plotting $\Delta
\rho_{xx}$ versus $1/(B\cos\theta)$ where $\theta$ is the angle between
$B$ and the $C_{3}$ axis. One can see that the oscillatory features are
essentially dependent on the perpendicular component of the magnetic
field. Also, as shown in Fig. 6(b), the angular dependences of all three
frequencies in the FT spectra are consistent with $1/\cos\theta$. These
results strongly suggest that the present SdH oscillations signify 2D
Fermi surface(s). We note that the SdH oscillations in this sample
disappeared after keeping the sample in ambient atmosphere for a week,
which suggests that the SdH oscillations were coming
from the surface. (This observation also suggests that the Al$_2$O$_3$
coverage, while useful for slowing the aging of the surface of
Bi$_2$Se$_3$, does not provide a perfect protection). Furthermore, one
can estimate the bulk mobility of this sample to be $\sim$40 cm$^{2}$/Vs
from the values of $\rho_{xx}$ and $R_{\rm H}$, and such a mobility is
too low to give rise to SdH oscillations of the bulk carriers below 14
T. All told, one can reasonably conclude that the observed SdH
oscillations are of the surface origin.

From the SdH-oscillation data, the Fermi wave vector $k_F$ can be
calculated via the Onsager relation $F$ = $(\hbar c/2\pi e)\pi
k_{F}^{2}$, yielding $k_{F}$ = 0.018, 0.029, and 0.032 \AA$^{-1}$ for
$F_{1}$, $F_{2}$, and $F_{3}$, respectively. These are of the same order
as the value $k_{F}$ = 0.031 \AA$^{-1}$ reported for the topological
surface state in $n$-type Bi$_{2-x}$Sb$_{x}$Se$_{3}$.\cite{Analytis1}
However, because of the multi-component nature of the oscillations that
leads to complicated waveforms, it is difficult to reliably extract the
cyclotron mass $m_{\rm c}$ nor the Dingle temperature for each component
using the Lifshitz-Kosevich theory.\cite{Shoenberg1984} This makes it
impossible to identify the origins of the three oscillation frequencies,
but possible reasons for the multiple components in the present case
include: (i) harmonics of a fundamental frequency are observed, (ii)
chemical potentials of the top and bottom surfaces are not identical and
give two frequencies associated with the topological surface states,
(iii) a trivial 2D electron gas \cite{Bianchi} created by the band
bending at the surface presents additional SdH oscillations. Note that,
in the case of the present sample used
for the detailed SdH measurements, the top surface was covered by
Al$_2$O$_3$ and the bottom surface was covered by the GE varnish, so the
conditions of the two surfaces were very different. To resolve the
origins of those multiple frequencies, an experiment involving the gate
control of the surface chemical potential to trace the energy dispersion
of each branch (as was done \cite{Sacepe} for exfoliated Bi$_2$Se$_3$)
would be desirable.

\subsection{Nonlinear $\rho_{yx}(B)$ behavior}

Although we could not extract the surface mobility from the SdH
oscillations in the present case, we can still estimate the relevant
parameters of the surface and the bulk transport channels by analyzing
the nonlinear $\rho_{yx}(B)$ behavior [Fig. 5(b)] with the simple
two-band model described in Ref. \onlinecite{BTS_Rapid}. The solid line
in Fig. 5(b) is the result of the two-band-model fitting, from which we
obtained the bulk electron density $n_{b}$ = 7$\times$10$^{17}$
cm$^{-3}$, the bulk mobility $\mu_{b}$ = 17 cm$^{2}$/Vs, the surface
electron density $n_{s}$ = 1.8$\times$10$^{12}$ cm$^{-2}$, and the
surface mobility $\mu_{s}$ = 1.2$\times$10$^{3}$ cm$^{2}$/Vs. The
observed frequencies of the SdH oscillations, 11, 27, and 33 T
correspond to the surface carrier densities of 2.5$\times$10$^{11}$,
6.7$\times$10$^{11}$, and 8.1$\times$10$^{11}$ cm$^{-2}$ in
spin-filtered surface states, respectively, and it is interesting that
the sum of these numbers, 1.7$\times$10$^{12}$ cm$^{-2}$, appears to be
consistent with the $n_{s}$ value obtained from the two-band analysis.
Also, it is assuring that the surface mobility $\mu_{s}$ obtained from
the two-band analysis, $\sim$1200 cm$^{2}$/Vs, is reasonably large and
is consistent with our observation of the SdH oscillations in moderate
magnetic fields.

In passing, the $\rho_{yx}(B)$ data of the sample A shown in Fig. 4(g)
can also be fitted with the two-band model. The result of the fitting,
shown by the solid line in Fig. 4(g), yields the bulk
electron density $n_{b}$ = 1.3$\times$10$^{18}$ cm$^{-3}$, the bulk
mobility $\mu_{b}$ = 15 cm$^{2}$/Vs, the surface electron density
$n_{s}$ = 2.0$\times$10$^{11}$ cm$^{-2}$, and the surface mobility
$\mu_{s}$ = 1.0$\times$10$^{4}$ cm$^{2}$/Vs for this sample.

\section{Conclusion}
In conclusion, we demonstrate that with tactful defect engineering one can
optimize the transport properties of the topological insulator
Bi$_{2}$Se$_{3}$ to show a high bulk resistivity and clear quantum
oscillations.
Specifically, by employing a Se-rich crystal-growth
condition we achieved the $p$-type state in Bi$_{2}$Se$_{3}$ by
Cd-doping for the first time; we then employed careful post annealing
to tune the Se vacancies and achieved a high level of compensation,
where the acceptors and donors nearly cancel each other and the sample
presents a high $\rho_{xx}$ value exceeding 0.5 $\Omega$cm at 1.8 K and
shows 2D SdH oscillations consisting of multiple components below 14 T.

\begin{acknowledgments}
We thank T. Minami for technical assistance.
This work was supported by JSPS (NEXT Program and KAKENHI 19674002),
MEXT (Innovative Area
``Topological Quantum Phenomena" KAKENHI 22103004), and AFOSR (AOARD
10-4103).
\end{acknowledgments}

\end{document}